\documentclass[12pt]{article}
\usepackage{makecell}
\usepackage{multirow}
\usepackage{graphicx}
\usepackage{caption}
\usepackage{amsmath,amsfonts,bm,amssymb}
\usepackage{natbib}
\usepackage{array,dcolumn,booktabs,float}
\usepackage{amsfonts}
\usepackage{amssymb}  
\usepackage{amsthm}  
\usepackage[FIGTOPCAP]{subfigure}
\usepackage{setspace}
\usepackage{verbatim}
\usepackage{JASA_manu}
\usepackage{color}
\usepackage{xcolor}
\usepackage{soul}
\usepackage{url}
\usepackage{hyperref}
\usepackage{qtree}
\usepackage{pdflscape}
\usepackage{vruler}

\usepackage{tikz}
\usetikzlibrary{fit,positioning,calc}

\newcommand{\R}{\mathbb{R}}

\newcommand{\I}{1{\hskip -2.5 pt}\hbox{I} }

\newcommand{\mockalph}[1]{}

\bibpunct{(}{)}{;}{a}{}{,}

\title{Bayesian non-asymptotic extreme value models for environmental data}
\author{
Enrico Zorzetto\thanks{Division of Earth and Ocean Sciences, Duke
University, Durham, USA$\,$ \tt enrico.zorzetto@duke.edu} $\,$ Antonio Canale\thanks{Dipartimento di Scienze Statistiche, Universit\`a degli Studi di Padova, Padova, Italy$\,$ \tt canale@stat.unipd.it}, $\,$ and 
Marco Marani\thanks{Dipartimento di Ingegneria Civile, Edile ed Ambientale, Universit\`a degli Studi di Padova,
 \tt marco.marani@unipd.it}
}

\begin{document}
\maketitle
\abstract{
Motivated by the analysis of extreme rainfall data, we introduce a general Bayesian hierarchical model for estimating the probability distribution of extreme values of intermittent random sequences, a common problem in geophysical and environmental science settings. The approach presented here relaxes the asymptotic assumption typical of the traditional extreme value (EV) theory, and accounts for the possible underlying variability in the distribution of event magnitudes and occurrences, which are described through a latent temporal process. 
Focusing on daily rainfall extremes, the structure of the proposed model lends itself to incorporating prior  geo-physical understanding of the rainfall process. By means of an extensive simulation study, we show that this methodology can significantly reduce estimation uncertainty with respect to Bayesian formulations of traditional asymptotic EV methods, particularly in the case of relatively small samples. The benefits of the approach are further illustrated with an application to a large data set of 479 long daily rainfall historical records from across the continental United States. By comparing measures of in-sample and out-of-sample predictive accuracy, we find that the model structure developed here, combined with the use of all available observations for inference, significantly improves robustness with respect to overfitting to the specific sample. 
{\center \textbf{Keywords: }}
Compound Distributions; Bayesian Hierarchical Models; Extreme Value Theory; Peaks Over Threshold; Extreme Rainfall.

\section{Introduction}\label{intro}

The quantitative modelling of extreme events is of paramount importance in several disciplines, such as water science, geology, engineering, and finance, to name a few. In these contexts extremes are often defined as the maximum values observed in each year, or, more in general, as {\it block maxima} (BM). This approach avoids (by neglecting them) having to explicitly tackle issues related to seasonality, and introduces a unit of time to define the frequency of occurrence of extremes over time scales of applicative interest. This traditional approach has proven very fruitful and has generated a large theoretical body related to the max-stability property of the Generalized Extreme Value (GEV) distribution \citep{fisher1928limiting, gnedenko1943distribution, von1936distribution, coles2001introduction}. An alternative modelling approach is based on defining extremes as exceedances over a high threshold, described through the theory developed by \citet{balkema1974residual} and \citet{pickands1975statistical}. Both approaches are asymptotic in nature. 

In the Block Maxima approach, GEV is the non-degenerate distribution obtained for block maxima, after proper normalization, in the limit of an infinite number of independent and identically distributed (i.i.d) events in each block \citep{fisher1928limiting, gnedenko1943distribution}, result later extended to the case of weak dependence structure \citep{leadbetter1983extremes, leadbetter2012extremes}. Based on the value of its shape parameter, often denoted as $\xi \in \R$, the GEV family includes three possible limiting distributions for the block maxima: a double exponential
(\emph{Gumbel}, or \emph{EV1}, for $\xi = 0$), a heavy-tailed (\emph{Fre\'chet}, or
\emph{EV2}, for $\xi > 0$), and an upper bounded (\emph{inverse Weibull} or
\emph{EV3}, for $\xi < 0$) distribution.

Conversely, in the Peaks Over Threshold (POT) framework, the Generalized Pareto Distribution (GPD) is derived as a model for excesses over threshold, in the limit of the threshold tending to the upper end point of the underlying random variables' support \citep{davison1984modelling, smith1984threshold, davison1990models}. This approach was later also extended to the case of dependent sequences \citep{leadbetter1983extremes, smith1992extremal, bortot1998models}. The GEV and GPD parametric models, respectively derived through the BM and POT frameworks, are deeply connected. In particular, by modelling the magnitude of threshold excesses with a GPD and their frequency of occurrence through a Poisson point process, again one obtains GEV as a model for the block maxima \citep{davison1990models, coles2001introduction}, with a parameter $\xi$ equal to the corresponding GPD shape parameter. For a comprehensive introduction, see \citet{coles2001introduction}, \citet{de2007extreme} and \citet{embrechts2013modelling}.

Threshold models generally lead to a more efficient use of the data compared to the BM approach. However, the selection of the threshold is a relevant issue in this case, and a contrast exists between the desire of including as much data as possible in the EV model, while at the same time satisfying the asymptotic assumption, which would require the adoption of a high threshold. Therefore in general the optimal threshold selection requires a tradeoff between bias and variance of the resulting estimator \citep{embrechts2013modelling}. Several techniques have been developed for informing this decision \citep[see][]{dupuis1999exceedances, coles2001introduction, embrechts2013modelling, wadsworth2012likelihood}.

The wide popularity enjoyed by approaches based on the GEV distribution led much of the extreme-value literature to focus on the block-maxima alone, or on few values above a high threshold, discarding and neglecting the 'ordinary values' from which these large events are extracted. In turn, this caused the widely accepted traditional Extreme Value Theory (EVT) 1) to be based on asymptotic results, to avoid the need of specifying details about the underlying distributions of the 'ordinary events', and 2) to focus only on few selected events, thereby 'wasting' most of the available information.

These issues have been receiving an increasing attention in recent times. Hydrological applications of EV models have shown that the number of yearly events is rarely sufficiently large for the asymptotic argument to hold \citep{koutsoyiannis2004statistics,marani2015metastatistical}. Moreover, for some parent distributions commonly used in a wide class of environmental applications, the actual extreme value distribution has been noticed to converge to its theoretical limiting form at a slow rate \citep{cook2004exact}. This is for example the case of the Weibull parent distribution, a parametric model widely adopted to describe several natural processes---such as  wind speeds \citep{harris2014parent} and rainfall accumulations  \citep{wilson2005fundamental}---or in economics \citep{laherrere1998stretched}. 



A more practical problem is related to the estimation of the GEV distribution shape parameter, $\xi$, which controls the nature of the tail of the distribution. 
When applied to precipitation data, maximum likelihood and L-moments estimates of $\xi$ from block-maxima
 and POT techniques can be markedly biased depending on the 
size of available samples, and this can lead to an underestimation of the probability of large extremes in the case of small samples \citep{koutsoyiannis2004statistics,
papalexiou2013battle, serinaldi2014rainfall}. This issue can be mitigated by use of sample statistics that are more efficient and robust than traditional ones \citep{hosking1987parameter}, or, following a Bayesian approach, by penalizing the likelihood function with  'Geophysical Prior' distributions for $\xi$ \citep{martins2000generalized, coles2003fully}. However, the limits, both conceptual and practical, of an approach that on the one hand heavily censors the data and, on the other, suffers by estimation bias and uncertainty, remain.
Another limitation of the traditional EVT which has been recently pointed out is related to the assumption of a single and invariable parent distribution \citep{marani2015metastatistical}. In fact, many phenomena display changes in the event magnitude generation process that are imperfectly known and predictable due to the complexity of the system. In these circumstances the assumption of a time-independent form of the parent distribution can be questionable. Examples of this type of issues can be found in many Earth-system processes and variables, such as rainfall intensity \citep{marani2015metastatistical, marra2018metastatistical}, flood magnitudes \citep{miniussi2020metastatistical}, wind speeds, and tropical storm intensities \citep{hosseini2020extreme}.
Overall, though mitigated by advanced estimation approaches, the above limitations can have wide implications in the many applications requiring the accurate estimation of large quantiles, i.e. quantiles characterized by return
times---average recurrence intervals---larger than the length of observed samples.

Recent contributions attempt to fill some of the gaps discussed above.
Some of these contributions have focused on including the entire parent distribution of events in EV modelling, by using mixture of distributions \citep{frigessi2002dynamic}, by extending a GPD model to the entire range of observed values while retaining a Pareto tail \citep{tancredi2006accounting, papastathopoulos2013extended, naveau2016modeling}, by combining splines with an algebraic tail decay \citep{huang2019estimating}, or by use of a parametric family of distributions to model the entire range of ordinary values \citep{marani2015metastatistical, joseph2019spatiotemporal}. 
The case of variable parent distribution has recently been tackled with the introduction of the Metastatistical Extreme Value Distribution (MEVD), a non-asymptotic extreme value approach in which a compound parametric distribution describes the entire range of ordinary values, with parameters varying across blocks \citep{marani2015metastatistical, zorzetto2016emergence, marra2018metastatistical, zorzetto2020extreme}. The main rationale behind the introduction of MEVD is describing the superposition of dynamics occurring over a wide range of time scales by use of compound distributions, i.e., by allowing the parameters of the distribution describing a \emph{fast} dynamics to vary on a separate, much slower time scale. 

Building upon the MEVD, here we introduce a Bayesian hierarchical model for extreme events which models the entire distribution of observed values, and explicitly incorporates the variability of their parent distribution across blocks. Latent variable models arise naturally in the Bayesian framework \citep{gelman2013bayesian} and in the context of extremes have been widely used to develop spatial models \citep{davison2012statistical, bracken2018bayesian} and to describe the temporal dependence of excesses over thresholds \citep{bortot2014latent, bortot2016latent}. Here we harness the flexibility of Bayesian hierarchical modelling to account for the low-frequency variability in the underlying physical processes generating the data observed in different blocks, and to connect this variability with the tail properties of their extreme value statistics.
The use of Bayesian methods to model extremes of environmental data is quite general and successful \citep{coles1996bayesian, coles2003fully, fawcett2018bayesian} and is particularly useful in the common case in which one has to rely on relatively short observational time series but has relevant and reliable expert prior information of the physical processes involved---as discussed in Section \ref{sec:priorelicitation}. 

The manuscript is organized as follows: In Section \ref{sec:structure} we  introduce the general structure of the hierarchical model and subsequently specialize it to the analysis of rainfall data with a focus on informative prior specifications. In Section \ref{sec:synthetic} the proposed formulation is empirically tested and compared to Bayesian implementations of standard extreme value models via a comprehensive simulation study. In Section  \ref{sec:rainfall} an application to a large  data set related to daily rainfall measured over the United States is described. The paper ends with a a final discussion.  The code used for the analysis is provided as a \texttt{R} package and is included in the online Supplementary Materials.

\section{A Hierarchical Bayesian extreme value model}
\label{sec:structure}

\subsection{Notation and general formulation}
\label{sec:model0}

The proposed Bayesian Hierarchical Model for Extreme Values (HMEV) is formulated by denoting as $n_{j}$ the number of events observed over the $j$-th \emph{block} of time ($j = 1, \dots, J$, with $J$ the number of blocks in the observed sample) and $x_{ij}$ the magnitude of the $i$-th event within the $j$-th block ($i = 1, \dots, n_j$). The magnitudes of the $n_j$ events occurring within a block are assumed to be realizations of independent and identically distributed (i.i.d.) random variables $X_{ij}$,
with common parametric cdf $F( \cdot; {\theta_j} )$. 
${\theta_j} \in \Theta$ is the possibly multivariate unknown parameter vector and $f(\cdot; \theta_j)$ the related probability density function.
%
Under this framework, the block maxima $Y_j = \max_i \left\{
X_{ij}\right\}$ have cdf
\begin{equation}
\zeta_j(y) = \mbox{Pr}(Y_{j} \leq  y) =  F(y; {\theta_j})^{n_j}.
\label{eq:cdfextreme}
\end{equation}
In the following we define a generative hierarchical model for the data at hand. A graphical representation of its structure is illustrated in Figure \ref{fig:specification}. We let $n_j$  be a realization of a random variable with  probability 
mass function (pmf) $p( n; \lambda )$, where $\lambda$ is an unknown vector of parameters.
We further assume that latent ${\theta_j}$'s exist that are i.i.d. realizations of a random variable with probability density function $g(\cdot; {\eta})$, where $\eta$ is an unknown vector of parameters. 
With the convention that the symbol $\sim$ means ``is a  realization of a random variable having pdf/pmf,'' we can write the following hierarchical model,
\begin{equation}
    n_j \mid \lambda \sim p(n_j; \lambda), \quad \quad \theta_j \mid \eta \sim g(\theta_j; {\eta}), \quad \quad x_{ij}\mid n_j, {\theta_j} \sim f(x_{ij}; {\theta_j}) \quad \text{for $i=1, \dots, n_j$}.
\label{eq:likelihood}
\end{equation}
Following a Bayesian approach, the hierarchical representation of the model is completed by eliciting suitable distributions, representing one's prior beliefs, for the unknown 
parameters ${\lambda}$ and ${\eta}$,
\begin{equation}
    {\lambda\mid \lambda_0} \sim \pi_{\lambda}( {\lambda}; {\lambda_0}), \qquad
    {\eta\mid \eta_0} \sim \pi_{\eta}( {\eta}; {\eta_0}).
\label{eq:prior}
\end{equation}
In equation \eqref{eq:prior} $\lambda_0$ and $\eta_0$ represent suitable prior hyperparameters. Comments and suggestions about their elicitation are reported in Section \ref{sec:priorelicitation}. 
Denoting as ${\boldsymbol{x}}$ the collection of all $x_{ij}$'s and as $\bf n$ the collection of all the $n_j$'s, we indicate with
$
\Pi(\eta, \lambda \mid \boldsymbol{x}, {\bf n}, \eta_0, \lambda_0) 
$
the posterior distributions of $(\eta ,\lambda) \in \Omega$.
\begin{figure}[t]
\centering
\begin{tikzpicture}
\tikzstyle{main}=[circle, minimum size = 7mm, thick, draw =black!80, node distance = 16mm]
\tikzstyle{connect}=[-latex, thick]
\tikzstyle{box}=[rectangle, draw=black!100]
\node[main, fill = white!100] (lambda0) [label=below:$\lambda_0$] { };
\node[main] (lambda) [right=of lambda0,label=below:$\lambda$] { };
\node[main, fill = black!10] (n) [right=of lambda,label=below:$n_j$] {};
\node[main] (eta0) [above=of lambda0,label=below:$\eta_0$] { };
\node[main] (eta) [right=of eta0,label=below:$\eta$] { };
\node[main] (theta) [right=of eta,label=below:$\theta_j$] { };
\node[main, fill = black!10] (x) [right=of n,label=below:$x_{ij}$] { };
\path (lambda0) edge [connect] (lambda)
(lambda) edge [connect] (n)
(n) edge [connect] (x)
(eta0) edge [connect] (eta)
(eta) edge [connect] (theta)
(theta) edge [connect] (x);
\node[rectangle, inner sep=10mm,draw=black!100, fit= (x) (theta), label=above:$j\in\{1\dots J\}$] {};
\node[rectangle, inner sep=10mm,draw=black!100, fit= (x) , label=above:$i\in\{1\dots n_j\}$] {};
\end{tikzpicture}

    \caption{Hierarchical structure of the  model described in equations \eqref{eq:likelihood}--\eqref{eq:prior}. Grey dots represent observed variables.}
    \label{fig:specification}
\end{figure}
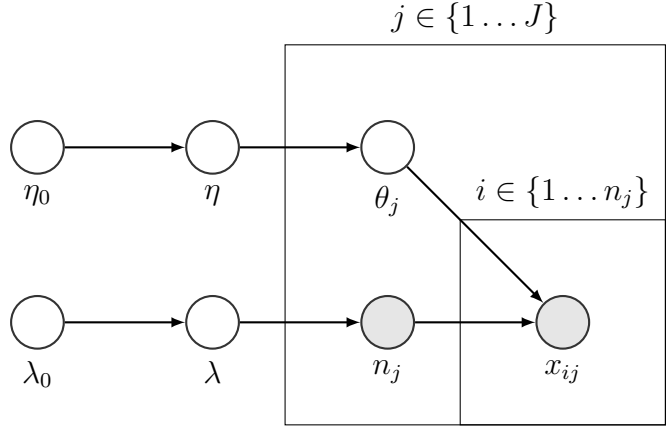




The main goal of extreme value analysis can be summarized in estimating the cdf in \eqref{eq:cdfextreme} or one of its functionals. This can be be done marginalizing out \eqref{eq:cdfextreme} with respect to the distributions of $\theta_j$ and $n_j$ \citep{marani2015metastatistical}, obtaining the following expression \eqref{eq:zeta}, where $h$ is function of the model's parameters $\lambda$ and $\eta$:
\begin{equation}
h(y ; \lambda, \eta) = \sum_{n = 0}^{N_t} \int_{\Theta} F(y ; \theta)^{n} g(\theta; \eta) p(n; \lambda)d\theta.
\label{eq:zeta}
\end{equation}
where $N_t$ is the maximum number of events in a block (e.g. $N_t=366$ days in the case of yearly blocks and daily observations of an environmental variable such as rainfall). A Bayesian estimator of \eqref{eq:zeta} can then be obtained by integration over the posterior distribution of the model parameters $\lambda$ and $\eta$:
\begin{equation}
\hat{\zeta}(y) = E[h(y ; \lambda, \eta) \vert x_{ij}, n_j] = \int_{\Omega} h(y;\lambda, \eta) \Pi(\eta, \lambda \mid \boldsymbol{x}, {\bf n}, \eta_0, \lambda_0) d\lambda d\eta.
\label{eq:zetaMC}
\end{equation}
Other functionals of interest such as the variance, or the probability intervals corresponding to given quantiles, can be calculated accordingly. As customary in extreme value analysis, for an event of given intensity $y$ we are interested in estimating the corresponding \emph{return time} $T_r$, or its \emph{average recurrence interval}, which is defined in terms of the cumulative distribution function as $\hat{T}_r(y) = \{1-\hat{\zeta}(y)\}^{-1}.$ Conversely, the return level $\hat{y}$ associated with a given non exceedance probability $p_{0}$, or return time $T_{r0} = 1/\left(1-p_0\right)$, is obtained as $\hat{y} =  \zeta^{-1} \left( 1 - 1/T_{r0}\right)$, where $\zeta^{-1}\left( \cdot \right)$ denotes the quantile function obtained by inverting the non exceedance probability function defined by eq. \eqref{eq:zetaMC}.

\subsection{A specific formulation of HMEV for modelling daily rainfall}
\label{sec:model}

We now discuss how the model structure presented above can be applied to modelling extreme values of environmental time series. Here we provide a specification of the HMEV for modelling the frequency of annual maxima daily rainfall accumulations, based on the general hierarchical structure outlined in Section \ref{sec:model0}. To this end, we need to specify parametric models for event magnitudes and occurrence, and elicit suitable prior distributions for their unknowns parameters. In this process, we seek to harness information on the physical processes generating the data and include it in the Bayesian pipeline. Several parametric families have been employed to model rainfall accumulations, including the exponential \citep{rodriguez1987some}, gamma \citep{stechmann2014first}, Weibull \citep{wilson2005fundamental}, lognormal and Pareto  \citep{papalexiou2013extreme}, or mixtures of Gaussian distributions \citep{li2013southeastern}. Generally, the choice of the model for a particular application is merely based on some goodness of fit assessment, without seeking a physical justification for the choice of the distribution. However, physical arguments have been provided suggesting the body of the daily rainfall distribution should follow a gamma distribution \citep{stechmann2014first, neelin2017global, martinez2019precipitation}, and suggesting its right tail should decay as a stretched exponential (i.e., Weibull) distribution \citep{wilson2005fundamental}. Since the focus of our work is on extreme values, we briefly review the latter  argument and show how physical insight can be incorporated into our Bayesian specification. \citet{wilson2005fundamental} noted that precipitation accumulations can be characterized as the product of three independent random variables, namely the average vertical air mass flux through a moist level, the air specific humidity, and the precipitation efficiency, i.e., the fraction of the vertical water vapor flux which is precipitated out as rainfall during each event. As these are all average quantities, it is assumed that, by the central limit theorem, their respective distributions can be approximated by Gaussians. By the theory of extreme deviations \citep{sornette2006critical, frisch1997extreme}, it can then be shown that, in the upper part of the distribution (i.e., for large enough rainfall accumulations), the product of a finite number $K$ of standard normal random variables is approximately a stretched exponential or Weibull distribution with a shape parameter equal to $2/K$, where $K=3$ is the number of variables in the present case. Therefore, not only this argument supports the choice of the stretched exponential distribution to model heavy rainfall accumulations, but additionally provides an indication on the value of its shape parameter. This argument provides valuable prior information to be exploited in our Bayesian hierarchical model. 
Consistently with this argument, we model the magnitudes of daily rainfall accumulations $x_{ij}$ in year $j$
with a 2-parameter Weibull distribution 
with parameter vector  ${\theta_j} = \left( \gamma_j, \delta_j\right)$ and pdf
\begin{equation}
    f_w(x; \gamma_j, \delta_j) = \frac{\gamma_j}{\delta_j} \left( \frac{x}{\delta_j} \right)^{\left( \gamma_j - 1 \right)}  \exp\left\{-\left(\frac{x}{\delta_j}\right)^{\gamma_j}\right\}
    \label{WEI}
\end{equation}
where  $\delta_j>0$ and $\gamma_j>0$ denote the  scale and shape parameters respectively.
To allow for the inter-block variability discussed in Section \ref{sec:model0},  we assume that the latent variables $\delta_j\sim g_\delta(\delta_j; \mu_{\delta}, \sigma_{\delta}) $ and $\gamma_j\sim g_\gamma(\gamma_j; \mu_{\gamma}, \sigma_{\gamma}) $ are independent and have Gumbel pdfs, a flexible yet parsimonious 2-paramater model allowing for possible asymmetry.
\begin{align}
&g_\delta(\delta_j; \mu_{\delta}, \sigma_{\delta}) = 
\frac{1}{\sigma_{\delta}} \exp{ \left\{ -    \frac{\delta_j - \mu_{\delta}}{\sigma_{\delta}} - \exp{ \left( - \frac{\delta_j - \mu_{\delta}}{\sigma_{\delta}} \right) }  \right\} } ,\\
&g_\gamma(\gamma_j; \mu_{\gamma}, \sigma_{\gamma}) = 
\frac{1}{\sigma_{\gamma}} \exp{ \left\{ -    \frac{\gamma_j - \mu_\gamma}{\sigma_{\gamma}} - \exp{ \left( - \frac{\gamma_j - \mu_{\gamma}}{\sigma_{\gamma}} \right) }  \right\} }  
\end{align}
Next, we need to specify $p(\cdot; \lambda)$ in equation \eqref{eq:likelihood}. It is well known that the rainfall process often tends to be overdispersed at the interannual time scale \citep{eastoe2010statistical}. This consideration would suggest a choice of $p(\cdot;\lambda)$ allowing a variance-to-mean ratio greater than one, to flexibly represent the possible presence of clustering. However, we show in the following that the distribution of $n_j$ chiefly affects the probability distribution of extreme events, \eqref{eq:zeta}, through its mean value only. To show this, let us rewrite \eqref{eq:zeta} in terms of the survival probability function $S(y ; \theta) = 1 - F(y ; \theta)$, 
\begin{equation}
h(y ; \lambda, \eta) = \sum_{n = 0}^{N_t} \int_{\Theta} [1- S(y ; \theta)]^{n} g(\theta; \eta) p(n; \lambda)d\theta,
\label{eq:zeta_napprox}
\end{equation}
by expanding $[1 - S(y;\theta)]^n$ in a Taylor series around zero $\left[ 1 -S(y ; \theta) \right]^{n}  = 1 - n S(y;\theta) +\mathcal{O} \left(S(y;\theta) \right) $, and by retaining only the linear term in the expansion---as justified for large values of $n$ and extreme quantiles (i.e., for $S(y \mid \theta) \to 0$)---one finds: 
\begin{align}
h(y ; \lambda, \eta) &\simeq 
\sum_{n = 0}^{N_t} p(n ; \lambda) \int_{\Theta} g		(\theta ; \eta) d\theta -
\sum_{n = 0}^{N_t} n  p(n ; \lambda) \int_{\Theta} S(y ; \theta) g(\theta ; \eta) d\theta \notag\\
& = 
1 - E_\lambda[n] \int_{\Theta} S(y ; \theta) g(\theta ; \eta) d\theta.
\label{eq:zeta_napprox2}
\end{align}
This expression depends on the distribution of $n_j$ only through its expected value conditional to the sample of observed $n_j$. We therefore argue for the adoption of a minimalistic model, the binomial distribution, with a success probability $\lambda \in (0,1)$ and number of trials $N_t$ equal to the block size (e.g., $N_t =366$ in our application to annual maximum daily rainfall). This rationale is also supported by practical applications of Poisson processes of extremes \citep{smith1989extreme} and of MEVD, showing that the specific distribution adopted for the $n_j$'s does not significantly affect the estimation of large extremes as long as the average is correctly reproduced \citep{marra2019simplified, hosseini2020extreme}

\subsection{Prior Elicitation} \label{sec:priorelicitation}

One of the main advantages of introducing a hierarchical model describing the entire distribution of daily rainfall accumulations is the possibility of eliciting priors directly on the underlying distribution of the observed ``ordinary'' events $x_{ij}$ and on the distribution of $n_j$, rather than on the distribution of block maxima. By doing so, in particular, we avoid the difficulty of prescribing a prior directly on the shape parameter $\xi$ of the annual maxima distribution, which is the main challenge in the inference on EV models, and to which it is difficult to attribute physical meaning. Studies at the global \citep{papalexiou2013battle} and continental scale \citep{papalexiou2018diagnostic} showed that the shape parameters of extreme value models can vary significantly in space and is particularly difficult to estimate reliably \citep[e.g.][]{coles2001introduction}, especially for small samples \citep{serinaldi2014rainfall}. However, here we argue that using the entire distribution of daily rainfall provides inferential advantages, and allows for the inclusion of additional physical insight on the process at hand.
For what concerns the specific parametric family for $\pi_\eta(\cdot; \eta_0)$, with $\eta = \left\{ \mu_{\delta}, \sigma_{\delta}, \mu_{\gamma}, \sigma_{\gamma} \right\}$, we opt for independent inverse gamma distributions but other choices of 2-parameters distributions such as gamma lead to a similar model flexibility and to qualitatively similar results. 
What is crucial is the specification of the values of the parameters of the above distributions according to our physical understanding of the precipitation process.  
Prior belief on the typical intensity of the events, $\mu_{\delta}$, is not difficult to obtain empirically for a given location as the climatological mean. 
Furthermore, the physical argument outlined in Section \ref{sec:model} enables us to assume a priori that the inverse gamma prior distribution for $\mu_\gamma$ is centered around $2/3$. Note that if additional physical insight is available on the types of storms characterizing the site of interest, or from similar sites, this prior elicitation could be further refined, e.g., based on studies of the value of $\gamma$ over large geographic areas \citep{papalexiou2018diagnostic}. 

For the latent Gumbel scale parameters $\sigma_{\delta}$ and $\sigma_{\gamma}$, quantifying the variability of the Weibull parameters between blocks, we also choose informative distributions with expectations equal to 25\% and 5\% of the respective location parameters ($\mu_{\delta}$ and $\mu_\gamma$). This choice reflects the notion that we expect significant variability in the scale parameter across years---here quantified as 25\% of its mean value ---but, conversely, we do not expect the shape parameter to vary as much, as its expected value should be more strongly constrained by the general physical nature of precipitation processes. Of course different precipitation types can occur in different proportions in different years, and, since we do not model these components explicitly, we should include their effect in possible variations of the scale parameters. Guided by these considerations, we choose a latent scale parameter for the variability of the Weibull shape parameter equal to 5\% of its prior expected value. 
Sometimes, information can be available on the relative frequency of different precipitation mechanisms, for example as obtained through satellite or radar measurements. In this case, the prior location value of $\sigma_{\delta}$ could be for example increased in settings characterized by higher inter-annual variability of the relative frequency of different precipitation types, as suggested in \citet{marra2019simplified}.
An independent weakly informative beta prior for the binomial rate parameter for $n_j$ concludes the prior elicitation. We found that eliciting an informative prior for $n_j$ is not as important as for the other parameters is the model, as (i) inference on the single-parameter distribution for $n_j$ is more robust than inference of the distribution of $x_{ij}$ even for very small sample sizes, and (ii) the HMEV estimates are primarily affected by the expected value of the $n_{j}$'s distribution rather than by its higher-order moments. The specific values of the prior parameters used in the in the remainder of the article are summarized in Table \emph{S1} in the online Supplementary Materials.

\subsection{Posterior computation and posterior predictive checks}
    \label{sec:computation}
Given the complex structure of the models described in previous sections, it is clear that an analytical expression for the posterior distribution of the parameters or for $\hat{\zeta}(y)$ in \eqref{eq:zetaMC} is not available and numerical procedures are needed. Here we chose to approximate the posterior distribution with Markov Chain Monte Carlo (MCMC) and specifically using a \emph{Hamiltonian
Monte Carlo} approach exploiting the flexibility of the Stan software \citep{carpenter2017stan}. 

The implementation of the hierarchical model and related prior described in Section \ref{sec:model} is trivial under Stan and is included in the Supplementary Materials as a standalone R package. In all the following examples, we run $n_c = 4$ parallel
chains, with $n_g = 2000$ iterations in each chain. We discard the first
half of each chain to account for the burn-in effect. 
The final sample on which we perform inference is therefore based on $B = n_c n_g/2 = 4{,}000$ draws.

Using MCMC we can make inference on any functional of the posterior distribution, calculating, at each iteration of the sampler, the current value of the functional of interests. For example, if the cumulative probability of block maxima approximating (\ref{eq:zeta}) is our target, one should compute at the generic iteration 
\begin{equation}
{\zeta}^{(b)}(y) = \frac{1}{M_g} \sum_{j=1}^{M_g} F(y; {\theta_j}^{(b)})^{n_j^{(b)}}
\label{eq:zeta_s}
\end{equation}
where $\theta_j^{(b)}$ and $n_j^{(b)}$ for $j = 1, \cdots, M_g$ are drawn from the related posterior predictive distributions for each block, and $M_g$ is a number of future blocks---$M_g = 50$ in our application. Therefore, the Monte Carlo approximation of the posterior  expectation (\ref{eq:zetaMC}) is 
\begin{equation}
\hat{\zeta}_{MC}(y) =\frac{1}{B} \sum_{b=1}^{B} {\zeta}^{(b)}(y).
\label{eq:zetaMC_s}
\end{equation}
Note that \eqref{eq:zeta_s} approximates the functional $h(z; \lambda, \eta)$ where $\lambda$ and $\eta$ are the parameters describing the inner level of the hierarchical model and the averaging operation in \eqref{eq:zeta_s} is performed on the values of $\theta_j$ and $n_j$. Conversely, \eqref{eq:zetaMC_s} is obtained by averaging over the $B$ draws from the posterior distribution thus  accounting for the posterior uncertainty of the $\lambda$ and $\eta$ parameters.

To assess whether the parametric assumptions of the proposed HMEV provide a good fit to the observed data, it is important to  perform posterior predictive checks \citep{gelman2013bayesian} comparing relevant quantities---such that $y_i$, $n_j$, or $x_{ij}$---with their corresponding posterior predictive densities.
Although the posterior predictive distributions are not analytically available, it is straightforward to simulate new data from them by leveraging the MCMC samples of the parameters and the hierarchical representation of the model reported in Figure \ref{fig:specification}. 
We recommend to focus on the distribution of block maxima, and,  given the interest in consistent estimates of the probability of large extremes, particularly on its right tail.

\section{Simulation study}

\subsection{Description}

To assess the empirical performance of the proposed HMEV model, and to compare it with standard alternative methods, we perform an extensive simulation study. Different synthetic data sets have been generated under four scenarios characterized by specific event magnitude distributions: Generalized Pareto (GP), Gamma (GAM), Weibull (WEI) with constant parameters in each block, and a dynamic Weibull model in which the variable scale and shape parameters in each block follow Gumbel distributions (WEI$_G$). 
 While the latter specification reflects the structure of the proposed hierarchical model, the other 3 scenarios represent model misspecifications and will be used to assess the rubustness of the proposed formulation to the specific distribution of event magnitudes. Common to all scenarios, the number of events in each block is drawn from a beta-binomial distribution with mean $\mu_n = 100$ events/block, variance equal to $\sigma_n^2 = 150$,  and $N_t = 366$ block size. This choice represents the case of overdispersion commonly observed in rainfall and other environmental time series \citep{eastoe2010statistical}. 
 Each of the $R_s = 100$ replicated data set consists of two independent time series of lengths $M_{train}$ and $M_{test}$ blocks, which are respectively used for training and testing the different EV models. Here we fix $M_{test} = 500$ yearly blocks, and train the different models focusing on sample size values of $M_{train} = 20$ and $50$ years, representative of many geophysical datasets.
 Table \emph{S2}, reported in the Supplementary Materials, describes the specific values of the parameters used to generate the synthetic data.

The competing methods used to benchmark HMEV are Bayesian implementations of the classical generalized extreme valued distribution (GEV) and peak over threshold (POT) Poisson point process models, whose details, including prior specifications, are reported in the Supplementary Materials. In order to perform a fair comparison, also these competing models are estimated under a Bayesian approach, using informative priors. In particular, for both models the prior distribution for the shape parameter is centered around the value $0.114$, determined from investigations of rainfall records at the global scale \citep{papalexiou2013extreme}, and has a standard deviation of $0.125$, yielding a distribution close to the Geophysical prior suggested by \citet{martins2000generalized}.

To evaluate the predictive accuracy of the different competing methods in estimating the true distribution of block maxima, we use different criteria measuring both the global goodness of fit and the uncertainty in estimating the probability of extreme events. 
The log pointwise predictive density (lppd) \citep{gelman2013bayesian} computed both for the in-sample data and for the out-of-sample data is often used as a measure of global performance of the models. An alternative measure is the logarithm of the pseudo-marginal likelihood (lpml), a convenient index that directly accounts---at no  additional computational cost---for a leave-one-out cross validation measure \citep{gelfand1994bayesian}. Notably, since the lpml approximates the expected log pointwise predictive density, the difference between the in-sample lppd and the lpml represents the number of effective parameters of a model  \citep[see e.g., ][]{vehtari2017practical} and thus will be used to quantify overfitting. 
Since the focus of this work is the right tail of the distribution of the block maxima, here we introduce an additional index that measures predictive performance for quantiles above a given non exceedance probability. To this end, we introduce the Fractional Square Error (FSE) 
\begin{equation}
FSE = \frac{1}{m_T} \sum_{j = 1}^{M_{x}}   \I_{({\tilde T}, \infty) }\left( T_{j} \right) \sqrt{ \frac{1}{B} \sum_{b = 1}^{B}
        \left( \frac{\zeta^{(b)^{-1}} \left( { p_j} \right) - y_j}{y_j} 
    \right)^2},
\label{eq:fse}
\end{equation}
where  $\zeta^{(b)^{-1}}(\cdot)$ refers to the quantile
    function  of the specific model at the $b$-th MCMC iteration, $\I_{A} \left( x \right) $ is the indicator function that equals 1 if $x$ belongs to $A$, and $T_j$ is the empirical return time of  $y_j$ defined as $T_{j} = (1-p_j)^{-1}$, with $p_j = \mbox{rank}(y_j)/(M_{x} + 1)$. $M_{x}$ is the length in blocks of the sample of annual maxima used to compute the FSE. In the in-sample and out-of-sample validation performed here, $M_x = M_{train}$ and $M_{x} = M_{test}$ respectively. The value $m_T$ represents the  number of observations in the test set with empitical return time equal to or larger than $\tilde T$, i.e. $m_T = \sum_{j = 1}^{M_{x}} \I_{({\tilde T}, \infty) } \left( T_{j} \right)$. Therefore, the FSE represents an average measure of a standardized distance between model-estimated quantiles and empirical quantiles for return times larger than $\tilde T$.
     In the following analysis we compute this measure for values of the return time larger than $\tilde T=2$ years, thus focusing on the range of exceedance probability of interest in many practical applications.
To separately assess the precision and the variability of  extreme value quantile estimates obtained from different models, we employ two additional measures, namely their average bias and the average width of the 90\%  posterior predictive credible intervals defined, respectively as
\begin{align}
        & b_q = \frac{1}{m_T}    \sum_{j = 1}^{M_{x}}  \I_{({\tilde T}, \infty) }\left( T_{j} \right) \frac{1}{B}  \sum_{b = 1}^{B} \left( \frac{\zeta^{(b)^{-1}} \left( { p_j} \right) - y_j}{y_j} 
    \right), \quad 
         \Delta_{q_{90}} = \frac{1}{m_T} \sum_{j = 1}^{M_{x}}  \I_{({\tilde T}, \infty) }\left( T_{j} \right) 
        \left( \hat{q}_{95} \left( p_j \right) - \hat{q}_{5}  \left( p_j \right) \right),
\label{eq:mwidth}
\end{align}
where the quantities $ \hat{q}_{95} \left( p_j \right) $  and $ \hat{q}_{5} \left( p_j \right)$ are the upper and lower bounds of the posterior credibility interval for the quantile $\zeta^{(b)^{-1}} \left( p_j \right)$ estimated taking the empirical quantiles over the  $B$ MCMC draws.

\subsection{Results}
\label{sec:synthetic}

\begin{figure}[t]
\centering
\includegraphics[width = 0.7\textwidth]{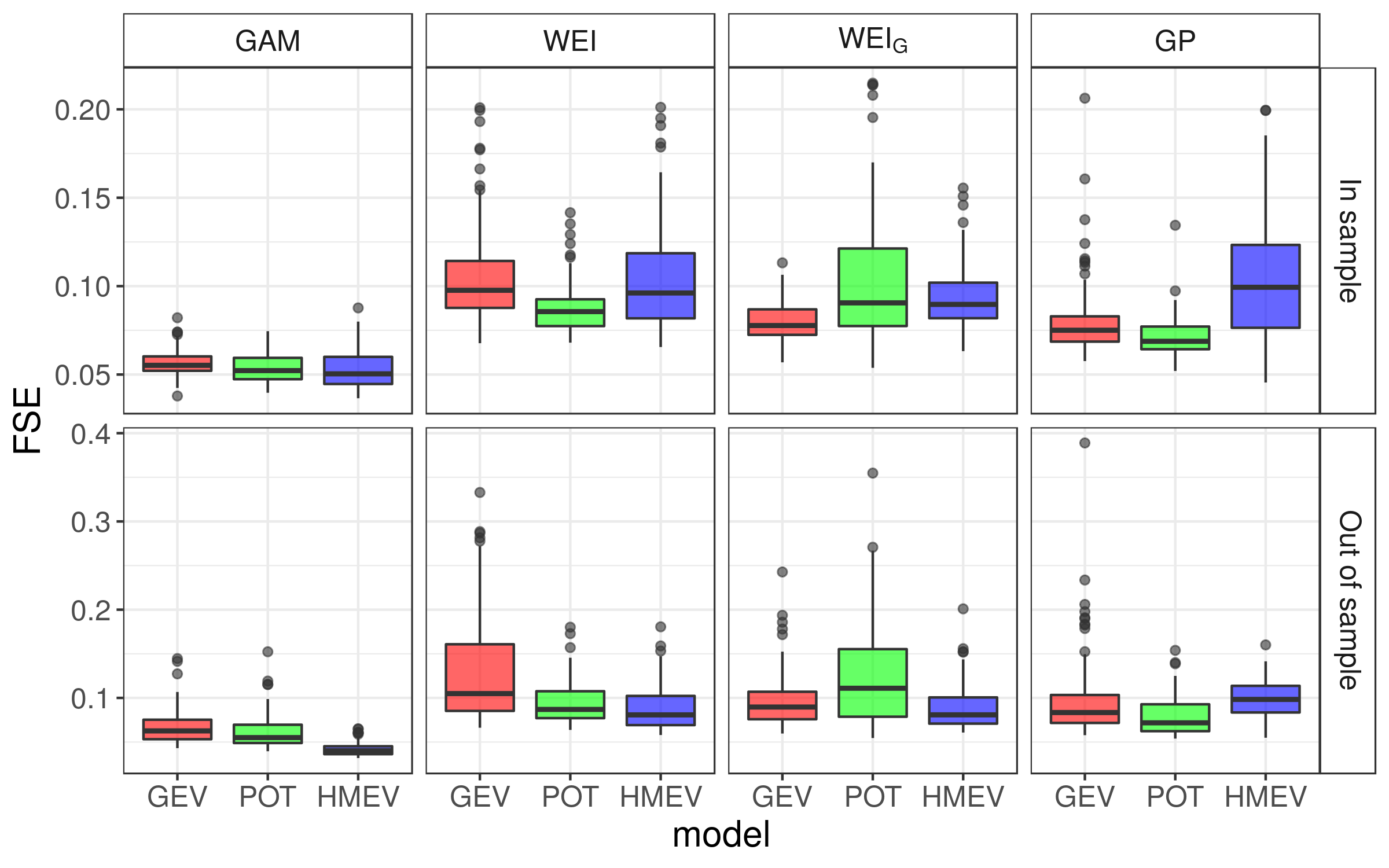}
\caption{Fractional square error computed for the 4 different model specifications for in-sample data (upper panels) and for out-of-sample data (lower panels), computed for a sample size of $50$ years.}
\label{fig:synth_fse}
\end{figure}

\begin{figure}
	\centering
	\subfigure[]{\includegraphics[width = 0.7\textwidth]{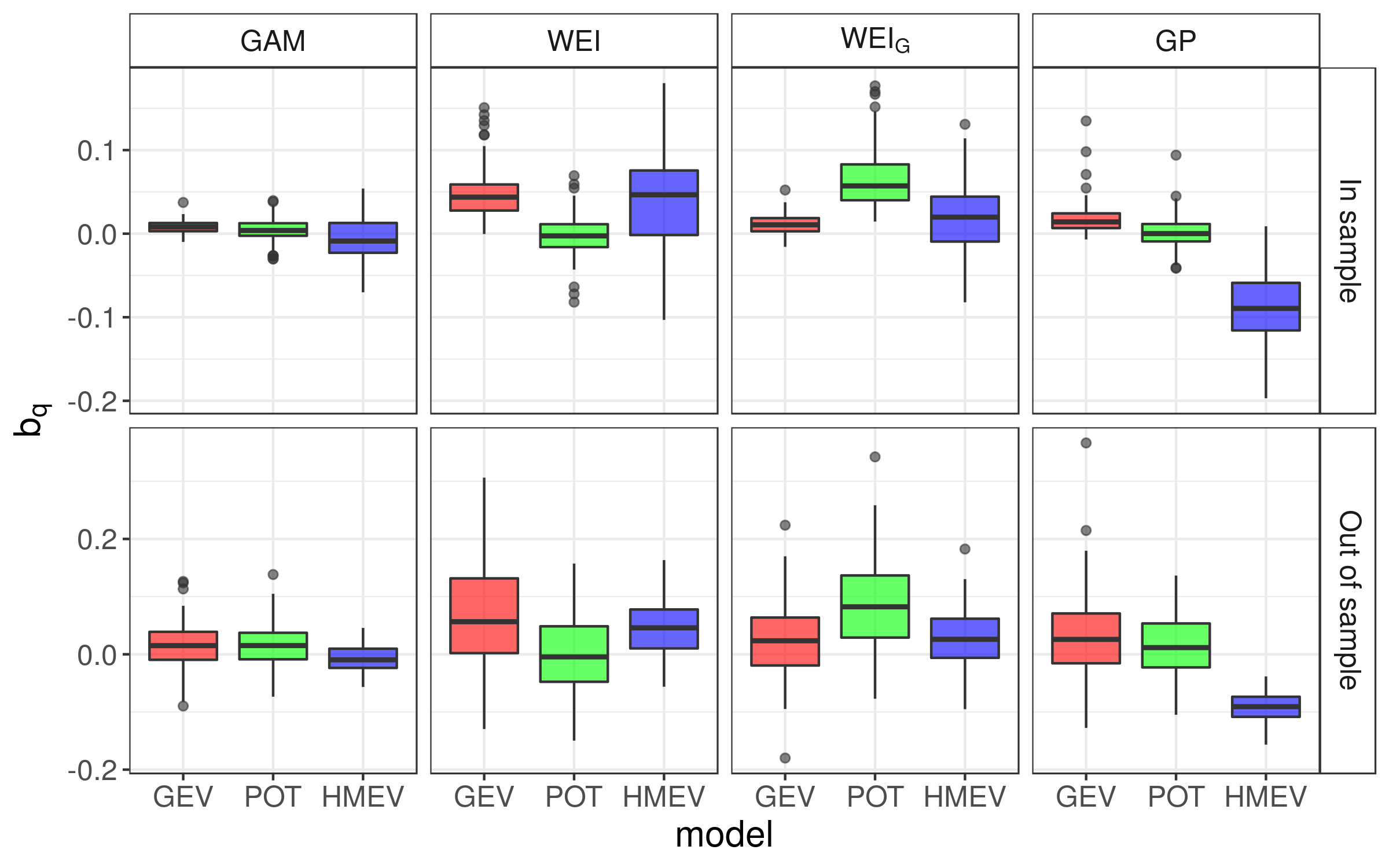}}
		\subfigure[]{\includegraphics[width = 0.7\textwidth]{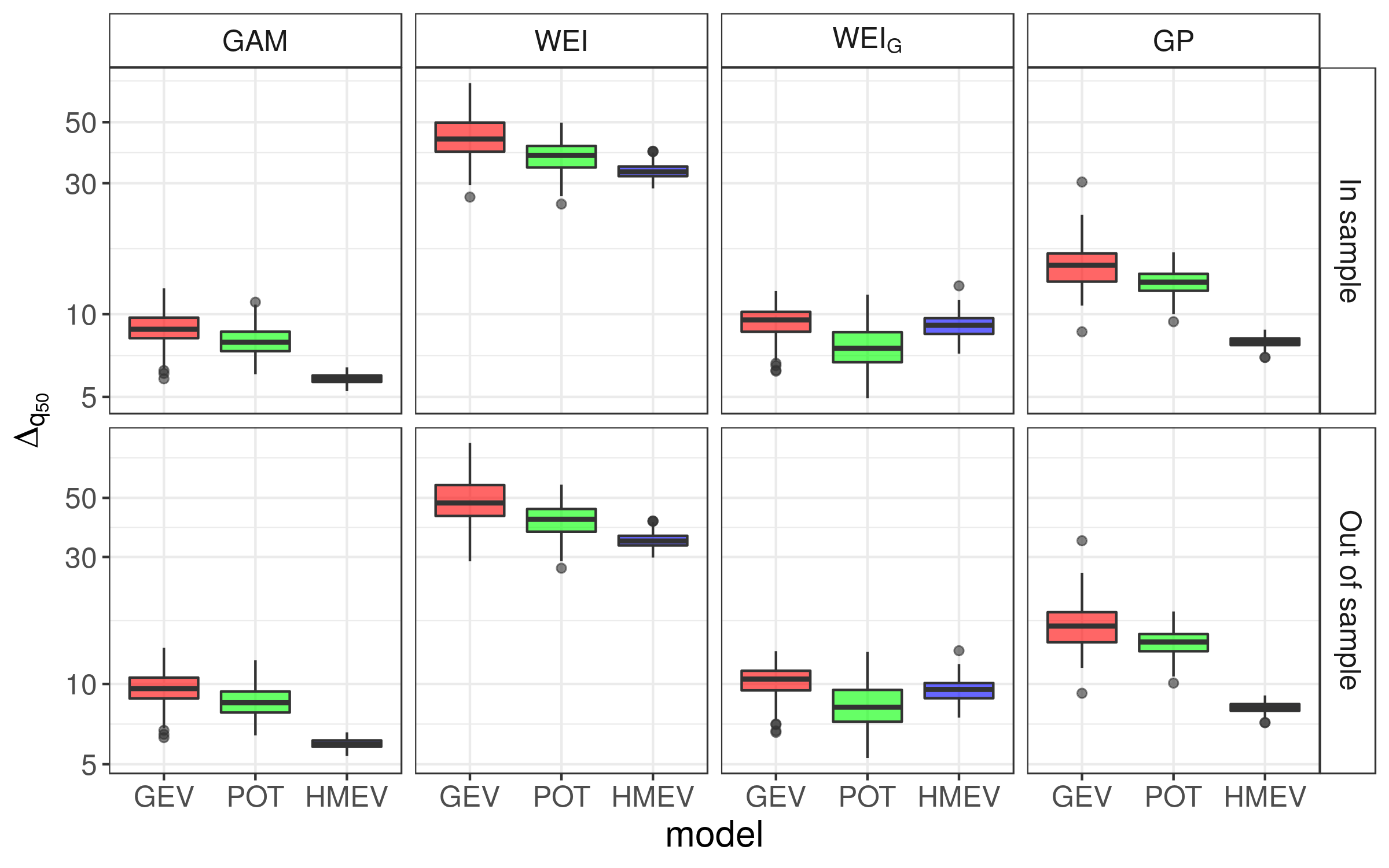}}
	\caption{Mean bias (a) and mean credibility interval width (b) for the 4 different model specifications for in-sample data (upper panels) and for out-of-sample data (lower panels), computed for a sample size of $50$ years.}
	\label{fig:synth_mwidth_and_mbias}
\end{figure} 

The results of the simulation study are illustrated in Figures \ref{fig:synth_fse}--\ref{fig:synth_enp}. Specifically, Figure \ref{fig:synth_fse} shows the empirical distribution of the FSE over the  $R_{s} = 100$ synthetic samples, training the model using $50$ years of simulated data.  The POT method appears to outperform the annual-maximum GEV in all cases examined, except in the case of WEI$_G$ specification, where arguably the inter-block variability of the $x_{ij}$ distribution determines a variable rate of threshold exceedance, as well as a variable distribution of the excess magnitudes over the fixed threshold. While exhibiting a generally higher FSE  for in-sample testing, HMEV cleary outperforms the competitors in the GAM, WEI, and WEI$_G$ scenarios in terms of out-of-sample performance. In the GP scenario, POT remains the best model even in the case of out-of-sample testing. 
To gain a deeper understanding of this general behavior, Figure \ref{fig:synth_mwidth_and_mbias}a reports the results of the two measures introduced in \eqref{eq:mwidth}.
Generally, the best performance for the bias appears to be  specification dependent, as is the case for the FSE, while for what concerns the width of the credibility interval, the HMEV is consistently the most efficient procedure, producing narrower credibility intervals. We note that the latent level temporal variability of the $\theta_j$ confer to HMEV a tail behavior which is intermediate between the lighter constant-parameter Weibull tail, and the Pareto model, as shown by the overestimation / underestimation of the posterior predictive quantiles in these two limiting cases. 
The bias of the different models does not appear to vary significantly from in-sample to out-of-sample testing, suggesting than the difference observed in the FSE is primarily controlled by the variability of the different estimates.

To visualize this global behavior, Figure \ref{fig:synth_example} shows a representative example of the performance of the methods. Specifically, it reports the quantile versus return time plots obtained for the different methods applied to a single dataset generated according to the WEI$_G$ specification, with the yearly number of events $n_j \sim Bin(\lambda)$, with $\lambda = 0.3$.
The results obtained for training datasets of 20 (panel a) and 50 (panel b) years show that  HMEV  yields quantile estimates characterized by narrower credibility interval and is characterized by a tail behavior which is lighter tailed compared to the other methods.
Note that both the GEV and POT models, despite the informative prior used, appear to be more sensitive to the largest observations in the training samples and tend to overestimate the true function. This behavior is expected given the limited length of the training samples used here (20 to 50 years of data), which are however representative of sample sizes commonly available in many applications in geophysics, engineering and environmental sciences. Representative plots for the remaining scenarios (model misspecification) are reported in the Supplementary Materials.
\begin{figure}[t]
\centering
\includegraphics[width = 0.7\textwidth]{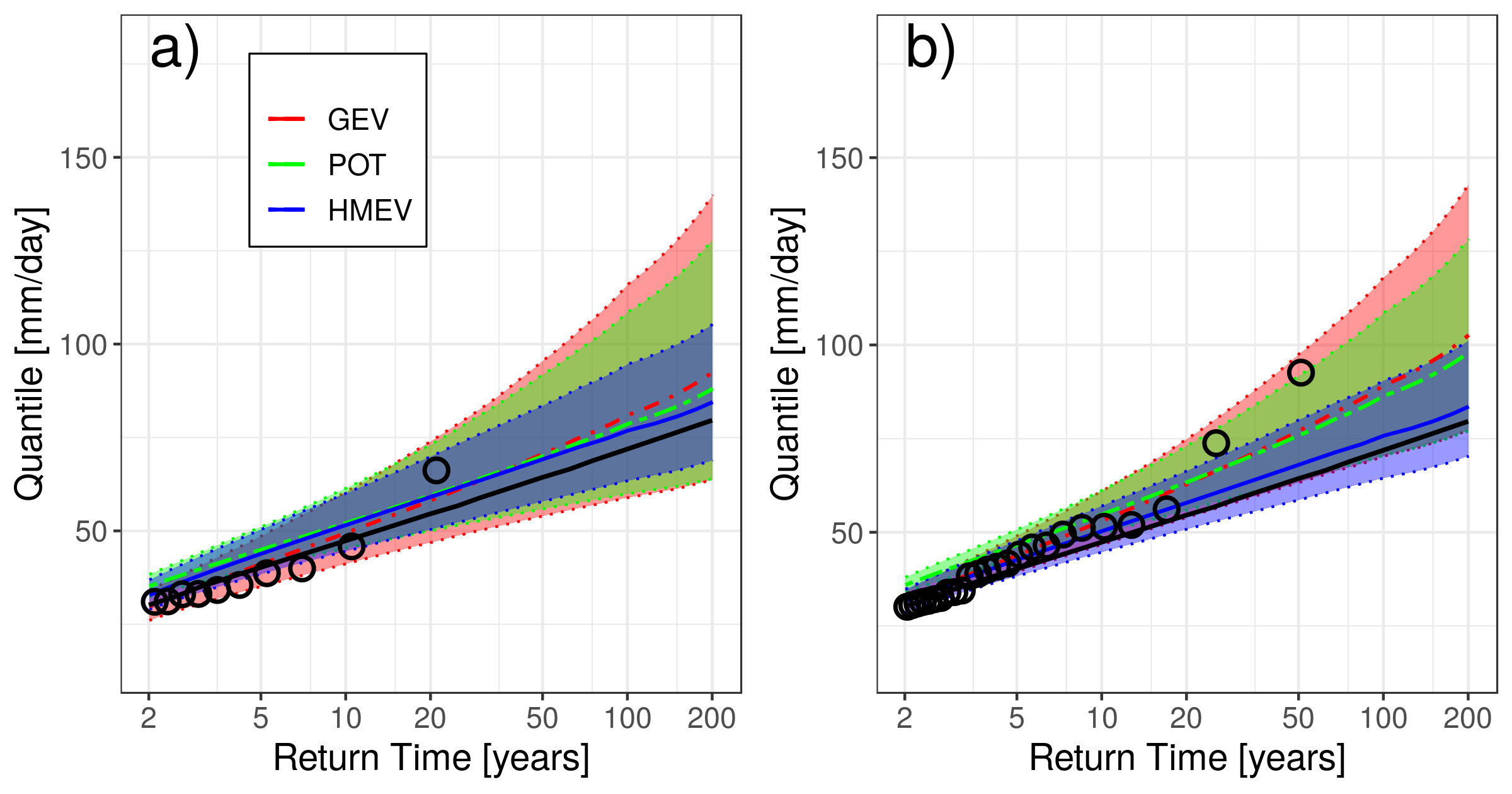}
\caption{Quantiles predicted by the GEV (red), POT (green), and HMEV (blue) models based on 		training sets of 20 years (a) and 50 years (b). Lines show the expected value of the quantile     for a given return time, while dashed lines represent the $5\%$-$95\%$ credibility intervals. Circles represent the observed return time of in-sample block maxima. The black lines report the quantiles computed from the true HMEV model.}
\label{fig:synth_example}
\end{figure}
Note that the in-sample and out-of-sample tests illustrated in Figures \ref{fig:synth_fse} and \ref{fig:synth_mwidth_and_mbias} are characterized by different sample sizes. Therefore, the absolute difference between in-sample and out-of-sample metrics is not directly interpretable as a measure of overfitting. Therefore, to better quantify overfitting we study the effective number of parameters in each model, estimated as the difference between the in-sample lppd and the log posterior marginal likelihood, shown in Figure \ref{fig:synth_enp}. HMEV displays a lower effective number of parameters in most  of the specifications considered,  suggesting that it is dramatically less prone to overfitting. While this behavior appears more markedly for two of the four data specifications ($WEI$ and $GAM$), it is worth noting that this advantage increases when considering sample sizes smaller than $M_{train} = 50$ years examined here, as shown in the Supplementary Materials.

\begin{figure}[t]
	\centering
	\includegraphics[width = 0.7\textwidth]{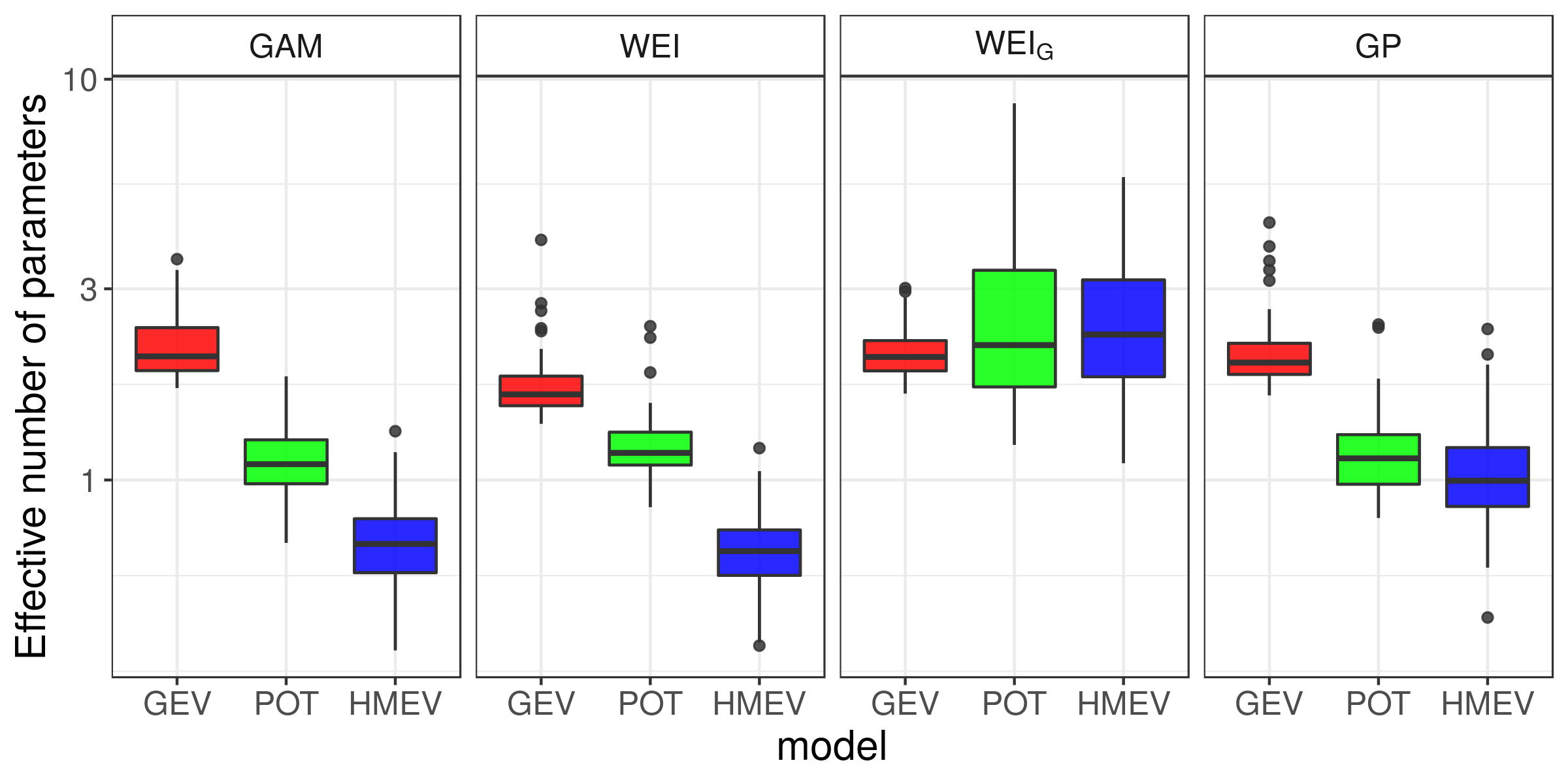}
	\caption{Effective number of parameters for the 4 different model specifications, evaluated for a sample size of $M_{train} = 50$ years of simulated data.}
	\label{fig:synth_enp}
\end{figure}

\section{Application to the United States Historical Climatological Network  data}
\label{sec:rainfall}

In this section we analyze a rich collection of daily rainfall time series extracted from the United States Historical Climatological Network (USHCN) data. The data are freely available from the National Centers for Environmental Information (NCEI) of the National Oceanic and Atmospheric Administration (NOAA) \citep{menne2012global, menne2012overview}. The USHCN data set consists of 1218 long daily rainfall records covering the Conterminous Unites States (CONUS), with a significant fraction of the available records being longer than 100 years. This caracteristic makes this dataset particularly useful for our purpose of assessing the performance of our method by using only a portion of the data for the model fit, keeping the remainder as out-of-sample validation data. Moreover, since the CONUS spans a range of different climatic regimes, this datasets allows us to test the robustness of the model structure adopted here to different climates and precipitation types, which are expected to impact both the distribution of the $x_{ij}$ and their arrival rate.
The records characterized by non-blank quality flag were removed from the analysis, as well as the years characterized by more than 30 daily missing observations. Therefore, for the subsequent analysis we select only stations with at least 100 years of record with enough non-missing, non-flagged observations, for a total of 479 stations.

\subsection{New York Central Park station analysis}

As a benchmark application we carefully discuss the results of the anlysis of the longest station in our data set, which was recorded in Central Park, New York City, from 1869 to 2018, for a total of 150 years of continuous observations (Station ID USW00094728). 
The entire series of daily event magnitudes as well as the 150 annual maxima values recorded at this station are reported in the top panels of Figure~\ref{fig:nycp_series}. 
\begin{figure}[t]
	\centering
	\includegraphics[width = 0.65\textwidth]{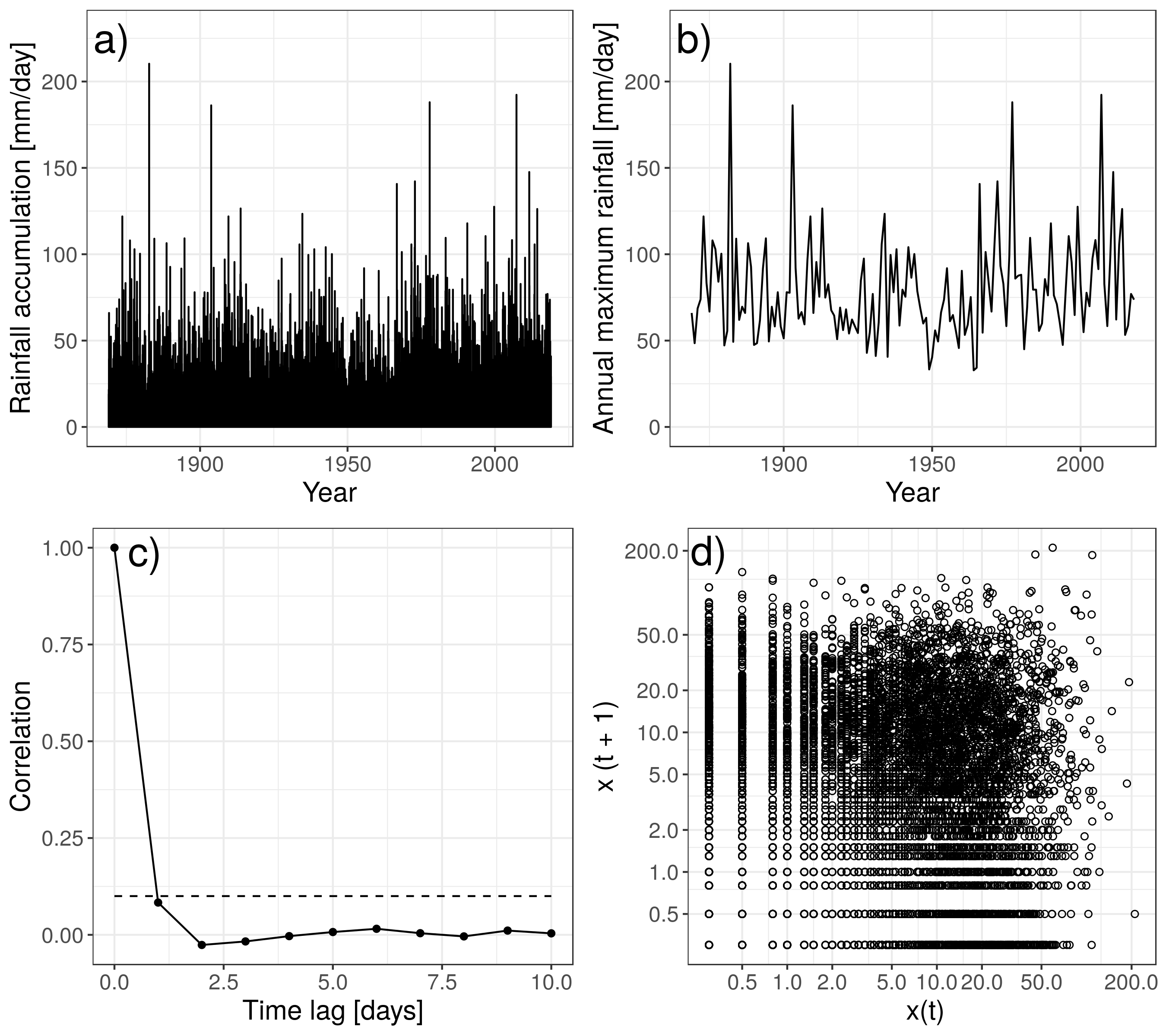}
	\caption{Rainfall time series measured at the New York Central Park (NYCP) station from 1969 to 2018 (Station ID USW00094728). (a), Time series of all daily rainfall accumulations, (b) annual maxima values only, (c) autocorrelation function of the daily rainfall accumulations, and (d) scatter plot of pairs of succeeding non-zero rainfall values.}
	\label{fig:nycp_series}
\end{figure}
Inspection of the autocorrelation function--- panel c) of  Figure \ref{fig:nycp_series}---suggests that the daily rainfall accumulations are not heavily correlated. 
 However, when dealing with rainfall accumulations at shorter time scales, or in different climatic conditions, serial dependence may need to be accounted for when applying extreme value models based on the i.i.d. assumption.
Hence, as commonly done in practice (see, e.g., \citet{cooley2007bayesian, marra2018metastatistical}), prior to model fitting we decluster the time series, by computing the autocorrelation of the daily magnitudes and determining the time lag $\tau_c$ in which the correlation decays below $c=0.1$. Then, each rainfall record is declustered by only keeping the largest accumulation value observed within a neighborhood of length $\tau_c$.

\begin{figure}[t]
	\centering
	\includegraphics[width = 0.7\textwidth]{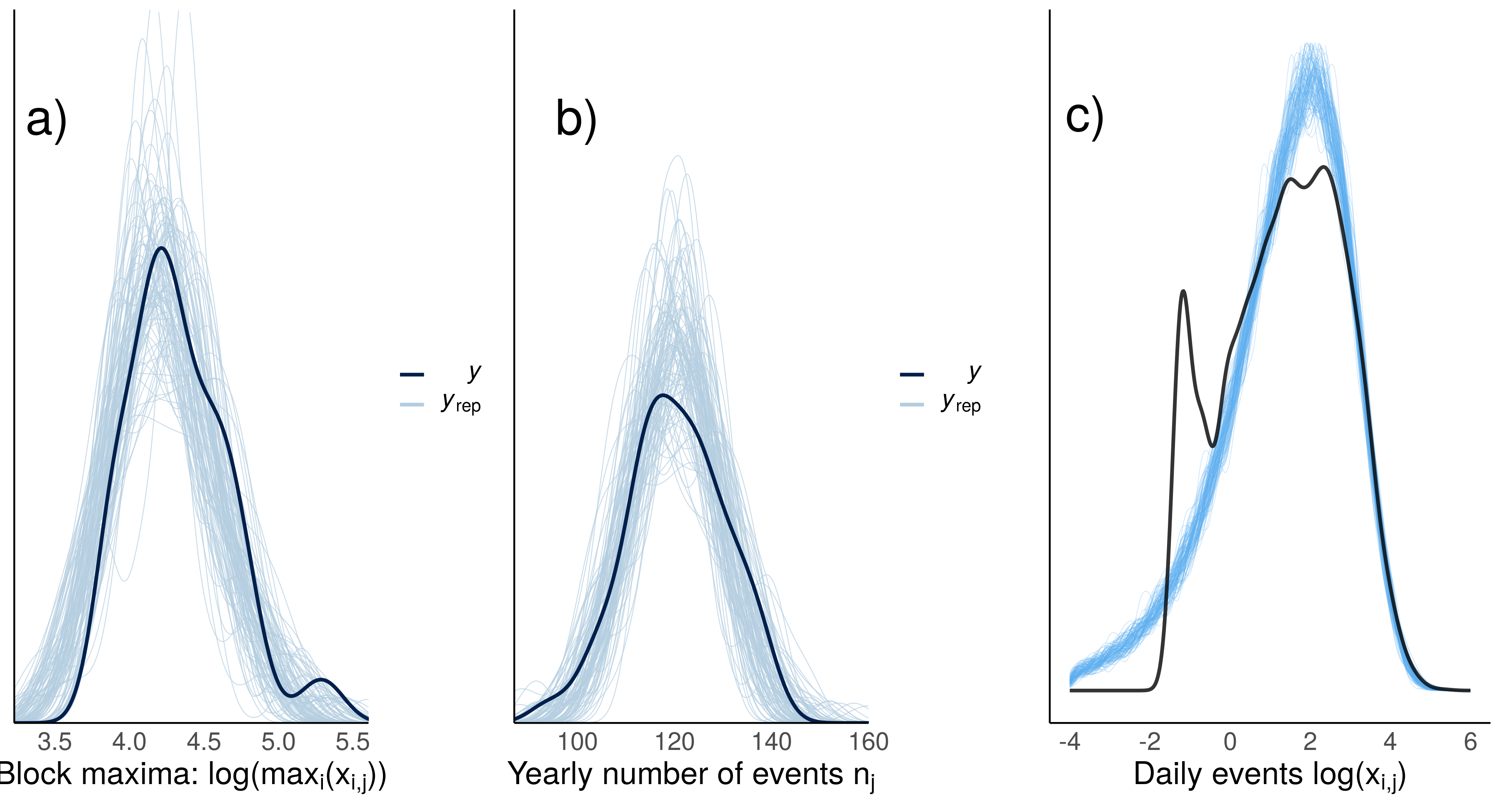}
	\caption{Posterior predictive distributions for the logarithm of the annual maximum daily rainfall accumulations (a), yearly number of events (b) and logarithm of non-zero daily rainfall events (c) computed by fitting HMEV to a 50-years sample extracted from the New York City time series. Black lines show the density of the observed values (obtained by kernel density estimation), while the light blue line show the kernel density estimates for 100 draws from the posterior predictive distributions.}
	\label{fig:ppdfs}
\end{figure}

After this selection of pseudo-independent events, data are analyzed following the approach described in Section \ref{sec:model}, with the same prior elicitation discussed in Section \ref{sec:priorelicitation}.
Examination of the posterior predictive distributions for the annual maxima, number of events, and daily rainfall magnitudes, reported in Figure \ref{fig:ppdfs}, shows that the pdfs of these variables are overall satisfactorily captured by HMEV. Note that a discrepancy appears for small values of daily rainfall magnitudes, where the censoring of actual values associated to the sensitivity of the instrument---0.3 $mm$ here---introduces a threshold in the observed accumulations---clearly visible in panel d) of Figure \ref{fig:nycp_series}. Despite this discrepancy for small magnitudes, the overall pdf of daily values, and in particular its right tail, appears to be satisfactorily captured by HMEV. 
The considerable length of this particular time series allows us to explore the sensitivity of extreme value estimates to the specific sample used to train the model. In Figure \ref{fig:nycp_quantiles} we compare extreme value quantiles obtained from the HMEV, GEV, and POT models trained on just the first 20 years or the first 50 years on record, respectively. Models estimates differ, with HMEV exhibiting---as previously observed from our simulation study---narrower credibility intervals with respect to POT and GEV models. HMEV predicts values slightly smaller than the Pareto model, but presents an overall good agreement with the empirical frequencies associated to the annual maxima extracted from the entire record (150 years of data). Interestingly, estimates from the GEV and POT models tend to fall between the frequencies computed from the training sets and those from the entire 150 year time series, showing their greater dependence on the specific training set used. This is shown even more clearly by the large differences, for GEV and POT estimates, between panel a) and panel b): when the length of the training set is increased such estimates significantly change, whereas HMEV estimates remain relatively insensitive to the increased available information.

\begin{figure}[t]
	\centering
	\includegraphics[width = 0.7\textwidth]{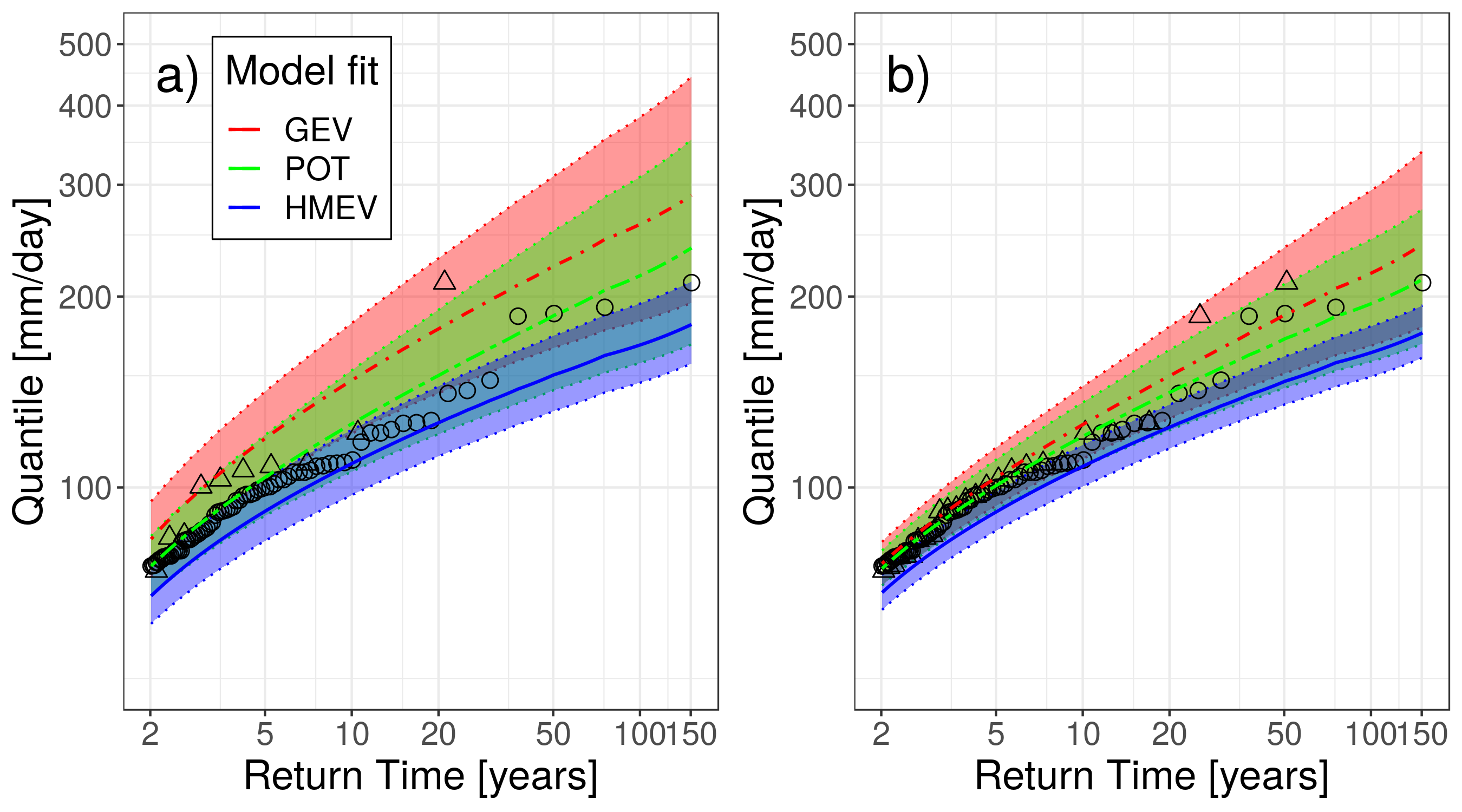}
	\caption{Extreme daily rainfall quantiles computed for the NYCP data set using for the fit only the first 20 (a) or 50 (b) years of the time series. Triangles represent the empirical cumulative frequency of data points in the training set, while black circles indicate the empirical frequencies computed from the entire 150-years time series. Predictions for the expected quantiles are indicated by red dashed line (Bayesian GEV), dashed green line (Bayesian POT), and continuous blue line (HMEV) with 5\% - 95\% credibility intervals reported as shaded areas for each model.}
	\label{fig:nycp_quantiles}
\end{figure}

\subsection{Full USHCN data analysis}

Building upon the insight gained from the simulation study, we now turn our attention to quantifying the predictive ability of different models using real observations from a larger set of stations, and repeating the analysis for different sample sizes in order to test the sensitivity of the different models to sample size, a well-known issue in applications of extreme value models. To this end, for each station in our sample of 479 USHCN datasets, we extract $M_{train}$ years to be used to train the extreme value model. Specifically, we repeat the analysis with $M_{train}$ equal to 10, 20, 30, 40, 50 years. In each case, we then randomly extract $50$ years of data from the remaining part of the time series to be used for independent validation. This procedure was repeated $R_{g} = 10$ times for each time series in the analysis reported here, thus producing a set of $4790$ sample points. For each of these cases, we investigate the effect of the specific model used and of sample size, employing the set of different performance metrics introduced in Section \ref{sec:synthetic}.

Figure \ref{fig:frac_best} shows the fraction of stations in which a specific competing model provides the best fit to the specific dataset. Examining this behavior for varying sample sizes, one can observe how HMEV becomes increasingly more competitive as the amount of available training data is decreased. As for simulation analyses, when one considers in-sample testing, POT most often is the best model. However, when out-of-sample performance is considered, the superior performance of the HMEV approach becomes clear. If one focuses on global measures of the probability distribution of estimation uncertainty for all yearly maxima, such as the lppd, the POT approach still seems superior for large training sample sizes. However, when the predictive uncertainty for extreme yearly maxima is examined (i.e. the FSE), arguably the main goal of extreme value analysis, the HMEV approach outperforms the other methods for all sample sizes. 
The difference between in-sample lpml and lppd, which provides a measure a model overfitting tendency, consistently depicts HMEV as the approach less prone to overfitting. See Figure \emph{S1} in the Supplementary Materials.

\begin{figure}[t]
	\centering
	\includegraphics[width = 0.8\textwidth]{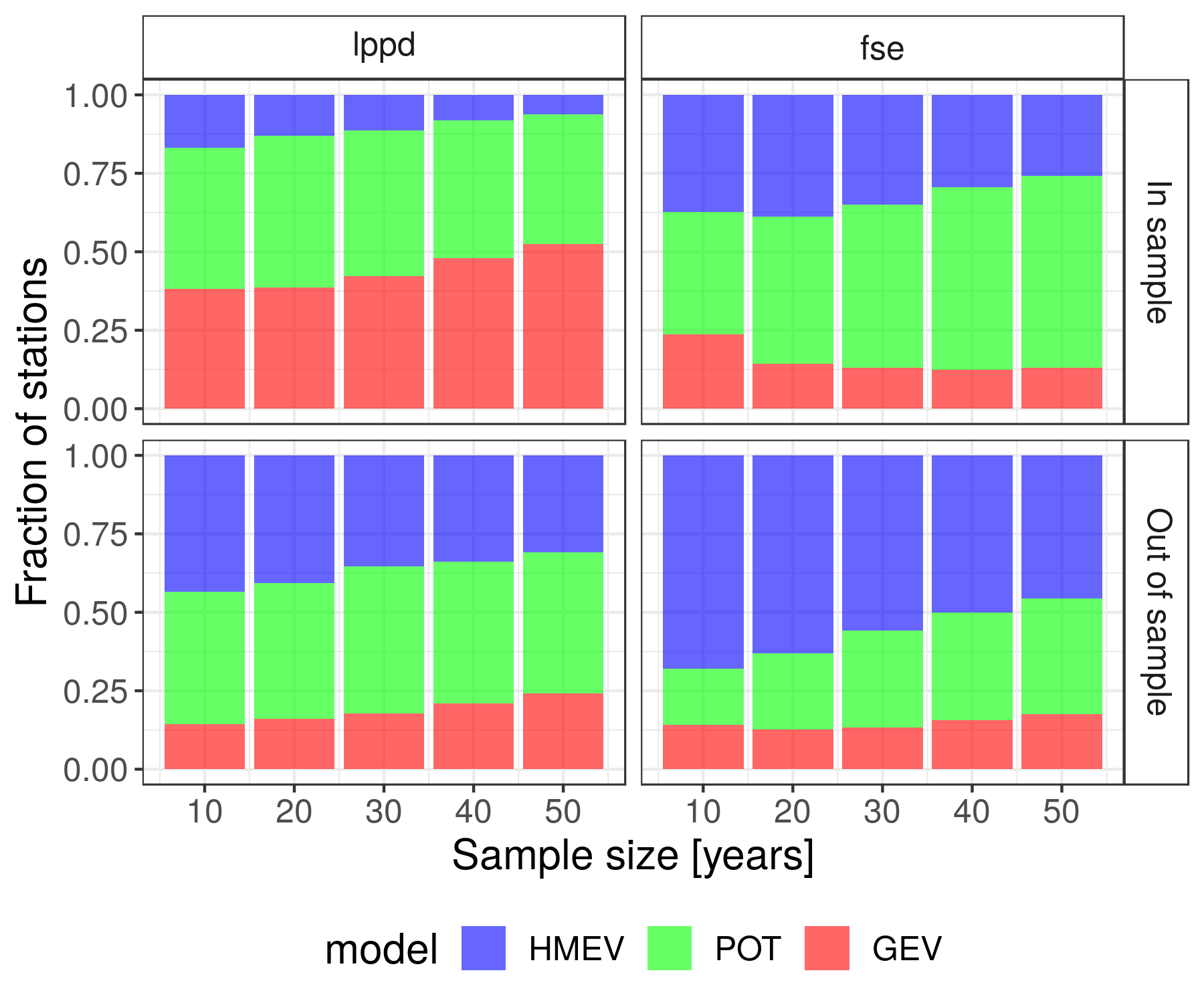}
	\caption{Fraction of cases in which each EV model exhibited the best estimation performance, evaluated with different metrics (FSE or lppd) using either in-sample or out-of-sample data.}
	\label{fig:frac_best}
\end{figure}

We also provide a spatially-explicit representation of model performances, by mapping, in Figures~\emph{S8} and \emph{S9}, reported in the Supplementary Materials, the best model for each station. 
The spatial distribution of the results overall appears to be consistent over the spatial domain of our study, even though some differences emerge between Western and Eastern USA. It is expected that, in specific locations, different precipitation regimes might produce distributions of daily rainfall which are not well captured by the stretched exponential model used in the present formulation of the HMEV. While producing location-specific models goes beyond the scope of the present study, the hierarchical structure of  HMEV can be flexibly adapted by using different parametric families for the parent distribution, while benefiting from the high predictive performance outlined in our analysis.

As a representative application of the HMEV method to the computation of extreme value rainfall quantiles, we report with different colors in Figure \ref{fig:quant_map} the magnitude of the 50-year daily rainfall event estimated for the set of stations analyzed here. For each station, following the analysis performed above, we randomly extract 50 years of record, repeating the procedure $R_{g} = 10$ times and averaging the results, so as to have a common same sample size for all records. As before, quantile estimates are obtained by numerically inverting the HMEV posterior predictive distribution, and computing the average quantile values over 4000 MCMC samples. This analysis provides a spatially explicit prediction for the 50-year event magnitude over the Continental United States, which as an example underlines the high quantile values corresponding to the South East, the Gulf coast, and the Pacific North West. The coherent probabilistic nature of the Bayesian HMEV can be exploited to assess the uncertainty of extreme value quantiles. For example, in Figure \ref{fig:quant_map} the width of the $90\%$ credibility intervals for a return time of $50$ years---normalized for the corresponding quantile---is proportional to the size of each dot. This relative measure of uncertainty appears to be larger in the Western USA, characterized by a drier climate and lower values of $50$-year quantiles.

\begin{figure}[h]
	\centering
	\includegraphics[width = 0.75\textwidth]{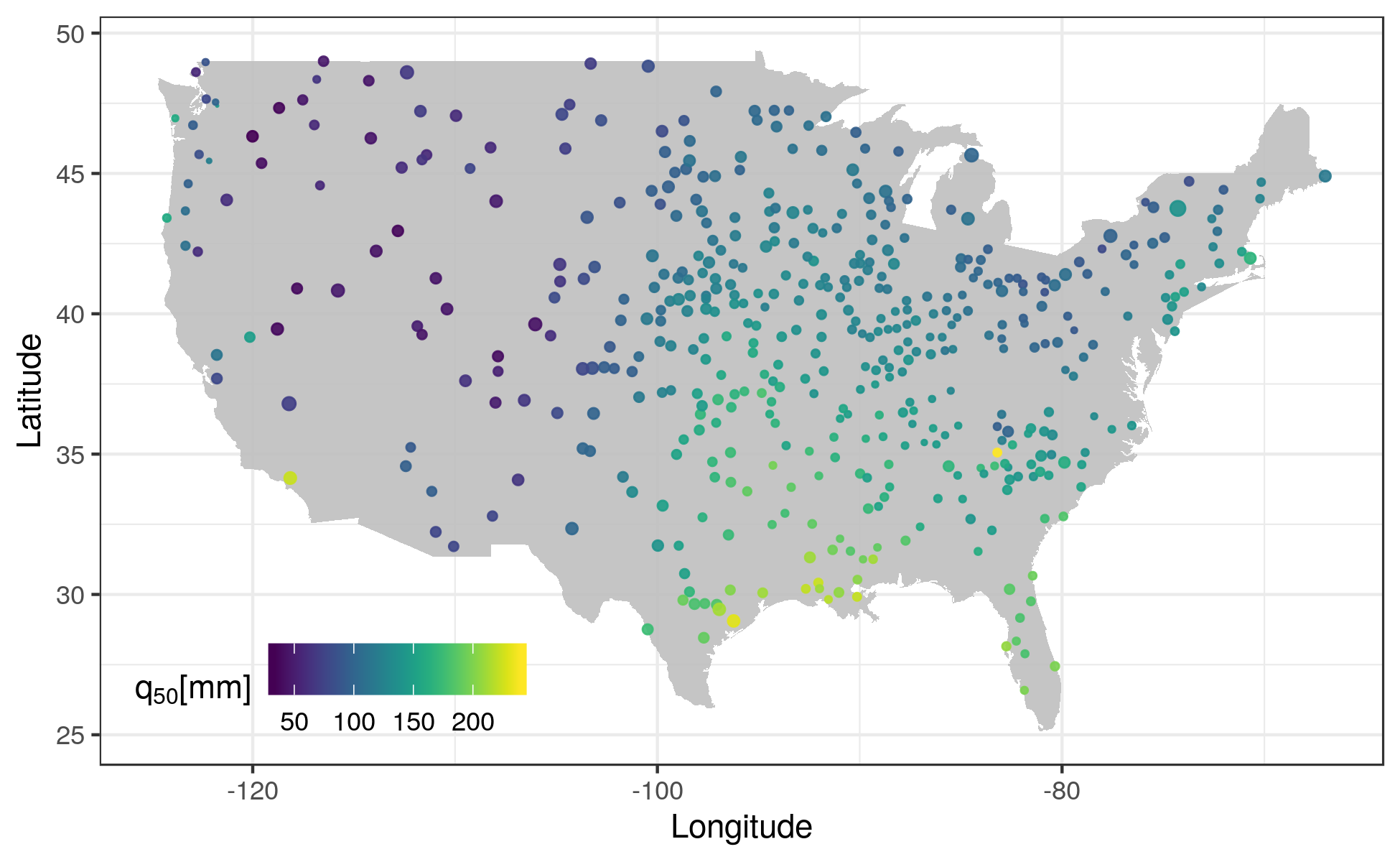}
		\caption{Spatial distribution of  the HMEV quantiles (color shading) and related normalized uncertainty (dot dimension) corresponding to an average recurrence interval of $T_r = 50$ years computed from 50-years samples extracted from the 479 USHCN stations included in the analysis. Normalized uncertainty computed as the ratio between the width of the $90\%$ credibility interval normalized and the posterior expected value of the quantile.}
		\label{fig:quant_map}
\end{figure}

\section{Discussion}
\label{sec:conclusion}

We introduced a Bayesian hierarchical model to make inference on extreme values of intermittent sequences, with underlying model parameters possibly varying over time. We  applied this approach both to synthetic and real data, testing its performance in estimating high quantiles, and provided a benchmark of its performance against some commonly used extreme value models. We found that the proposed approach can reduce uncertainty in extreme value estimation with respect to Bayesian formulations of other EV models. We attribute this behavior to the increased amount of observational information used in HMEV, and to the ability to leverage available information regarding the parent distribution describing the underlying physical process. This advantage becomes crucially important for short observational time series, and especially for large extremes in the right-most part of the distributional tail.
Our findings show that, when the underlying process generating the observations $x_{ij}$ is well approximated by a parametric model---such as the Weibull distribution adopted here--- use of an asymptotic extremal model leads to the loss of a large amount of information and to the subsequent inflation of the posterior uncertainty. 
While the ability of the proposed model to describe the tail of different processes appears to be dependent on the specific marginal distribution of $x_{ij}$, the model structure introduced here exhibits narrower posterior predictive intervals and a lower effective number of parameters when compared to other widely used extreme value models that do not attempt to account for the entire parent distribution. HMEV quantile estimates, in fact, exhibit reduced uncertainty even when the (synthetic) data being analyzed is not generated by the specific parent distribution of ordinary values chosen in the HMEV formulation. Therefore, a clear advantage in applying the HMEV methodology is that posterior predictive tests can be employed to check in-sample goodness-of-fit, and overfitting is minimized. 
In addition to these advantages, HMEV, based as it is on the specification of a distribution for all observations, is also amenable to possible extensions and generalizations. For example, at locations where different event-generating mechanisms are present \citep{li2013southeastern, marra2019simplified}, one can quite naturally adopt more complex specifications for the distribution of event magnitudes, such as mixtures of parametric distributions, as was done in some  MEVD formulations \citep{marra2018metastatistical, miniussianalyses}. 
Finally, the HMEV framework also naturally lends itself to extensions aimed at including possible systematic changes in the probability distributions of ordinary values, e.g. associated with trends in low-order moments derived from observations, climate model projections, or from physical principles that may provide insight into future rainfall characteristic magnitudes (e.g. Clausius-Clapeyron scaling of atmospheric water-holding capacity \citep{allan2008atmospheric}).

\section*{Acknowledgements}

The first author acknowledges support from the National Aeronautics and Space Administration through the NESSF program, fellowship 19-EARTH19R-27. The second author is supported by the University of Padova under the STARS Grant. Marco Marani acknowledges support by Provveditorato for the Public Works of Veneto, Trentino Alto Adige, and Friuli Venezia Giulia, through the concessionary of State Consorzio Venezia Nuova and coordinated by CORILA.

\appendix

\bibliographystyle{asa}
\bibliography{hmev_bib}

\end{document}